\documentclass[12pt]{iopart}

\usepackage{graphicx}
\usepackage{hyperref}
\usepackage{enumitem}
\usepackage[symbol]{footmisc}

\hypersetup{
    colorlinks=true,
    linkcolor=blue,
    filecolor=magenta,      
    urlcolor=cyan,
    pdftitle={Overleaf Example},
    pdfpagemode=FullScreen,
    }
\begin{document}

\title{Estimation of $\pi$ via experiment}
\setcounter{footnote}{0}       
\author{Keiko I. Nagao$^{a}$\footnote{Corresponding author}}
\address{$^a$Okayama University of Science, 1-1 Ridaicho Kita-ku, Okayama, Okayama 700-0005 Japan}
\ead{nagao@ous.ac.jp}

\author{Yuga Sakano$^{a, b}$}
\address{$^b$Hiroshima University, Higashi-Hiroshima, Hiroshima 739-8526, Japan}
\author{Takashi Shinohara$^{a}$}
\author{Yuji Matsuda$^{c}$}
\address{$^c$National Institute of Technology, Niihama College, 7-1 Yakumocho, Niihama, Ehime 792-8580 Japan}

\author{Hisashi Takami$^{a}$}

\vspace{10pt}
%\begin{indented}
%\item[]February 2025
%\end{indented}

\begin{abstract}
In this study, we conducted an experiment to estimate $\pi$ using body-to-body and body-to-wall collisions. 
By geometrically analyzing the system’s motion, we first review how the collision count corresponds to the digits of $\pi$.
This method utilizes the property that the number of collisions corresponds to $\pi$ to the $n$-th decimal place by setting the mass ratio of bodies to $1:100^n$ under ideal conditions. In particular, when the mass ratio is $1:100$ — which is the case we tested experimentally — the number of collisions is 31, and $\pi$ to the tenths decimal place (3.1) can be derived.
In the experiments, a suspended apparatus was developed to minimize energy losses such as friction and air resistance. We also devised the shape and material of the colliding bodies' surface and the characteristics of the suspension string, aiming for measurements under stable conditions. 
Based on the experimental results, we reproduced the number of collisions consistent with the theoretical values and confirmed that estimating $\pi$ to the tenths decimal place is possible under realistic conditions.
%We experimentally measured $\pi$ using object-to-object and object-to-wall collisions, achieving 3.1 (one decimal place) with a mass ratio of 1:100. To minimize energy loss, we designed a suspended apparatus and optimized the collision conditions. Our results validate the feasibility of measuring $\pi$ experimentally under realistic conditions.
\end{abstract}

%
% Uncomment for keywords
%\vspace{2pc}
%\noindent{\it Keywords}: XXXXXX, YYYYYYYY, ZZZZZZZZZ
%
% Uncomment for Submitted to journal title message
%\submitto{\JPA}
%
% Uncomment if a separate title page is required
%\maketitle
% 
% For two-column output uncomment the next line and choose [10pt] rather than [12pt] in the \documentclass declaration
%\ioptwocol
%

\section{Introduction: Estimating $\pi$ by collision of two point masses and a wall}
\label{sec:intro}

The existence of $\pi$ has been known since ancient times. 
Particularly, in modern times, attempts have been made to precisely determine the value of $\pi$ to the largest decimal place using high-performance computers.
Recent studies have also shown that $\pi$ can be obtained experimentally from the number of collisions between two masses and a wall, gaining significant attention.
When two masses with a mass ratio of $1:100^n\, (n = 0, 1, 2, 3, \cdots)$ and a wall collide, the number of collisions corresponds to $\pi$ to the $n$-th decimal place \cite{Galperin, Sanderson, Brown:2019jvs, Aretxabaleta, Cai:2022jcv}. For example, if the mass ratio is $1:100^3$, the number of collisions between two point masses and a wall is 3141, and $\pi=3.141$ can be derived. However, the number of collisions coincides with $\pi$ only in the ideal situation where energy loss due to friction, air resistance, etc. can be ignored. It is not obvious that $\pi$ can be accurately obtained in actual experiments.
In this study, several devices were built and tested, and it was confirmed that $\pi$ up to 3.1 can be obtained when the mass ratio is $1:100$. This paper is organized as follows. 
In Section~\ref{sec:theory}, we provide a overview of the theoretical background for this experiment.
Section \ref{sec:experiment-setup} describes the problems in conducting these experiments based on the preparatory experiments and our experimental apparatuses. 
In Section \ref{sec:experiment-result}, we measure the number of collisions using the experimental apparatus and discuss the results. Section \ref{sec:conclusion} provides a summary and discussion.

%-----------------------------------------------------------------
\section{Theoretical Background}
\label{sec:theory}

\subsection{Velocity–Space Geometry and Conservation Laws}
\begin{figure}[thbp]
  \centering
  \begin{minipage}[c]{0.35\textwidth}
    \includegraphics[width=\linewidth]{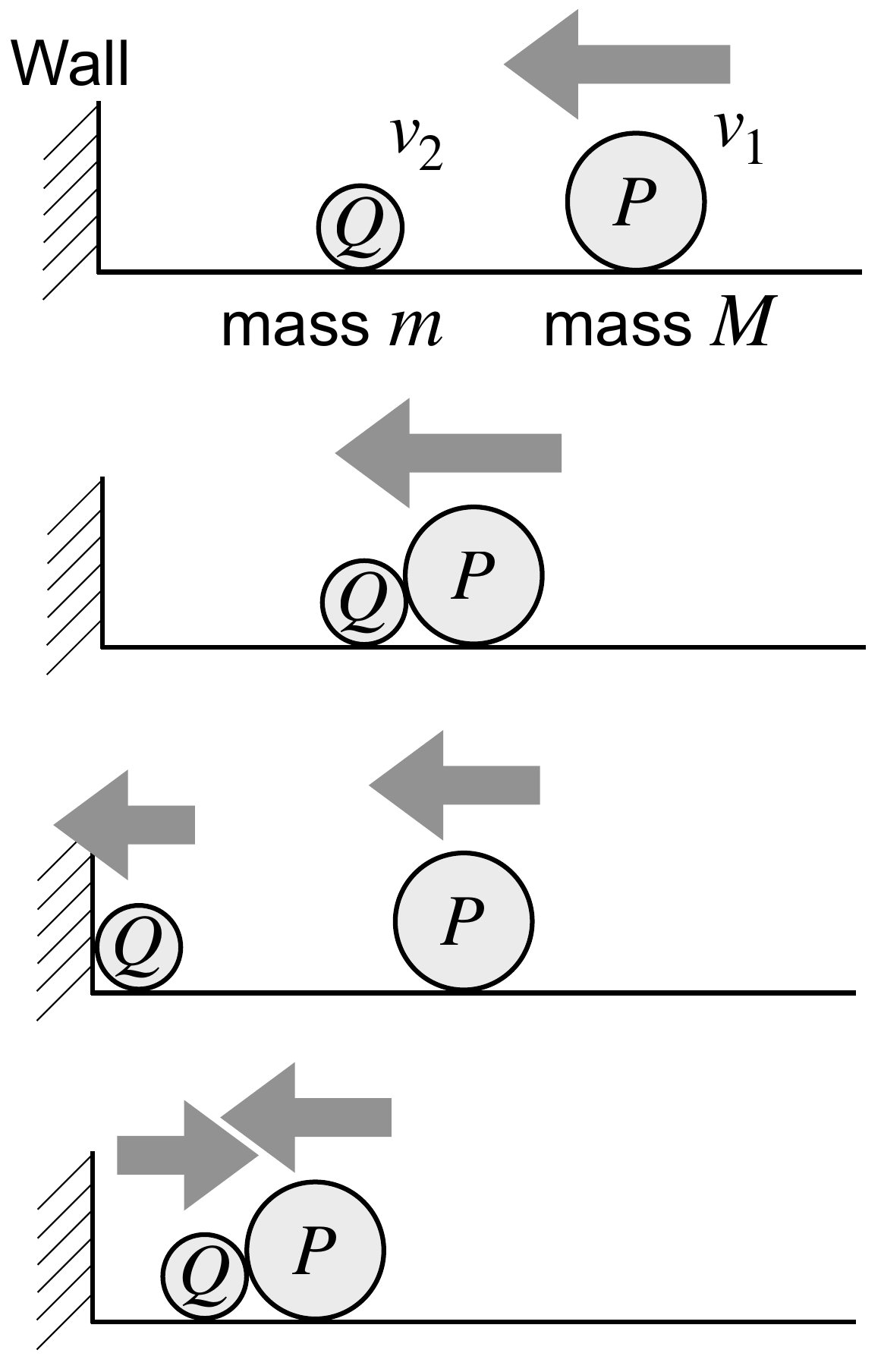}
  \end{minipage}%
  \hfill
  \begin{minipage}[c]{0.62\textwidth}
    \small
    \textbf{Collision Sequence}\\[0.3em]
    \textbf{(0)} Initial state.
    Body $P$ of mass $M$ moves toward stationary body $Q$ of mass $m$.\\[2.5em]
    \textbf{(1)} After $P$ hits $Q$, 
    a portion of $P$'s momentum is transferred to $Q$. Body $Q$ moves toward the wall.\\[2.em]
    \textbf{(2)} After $Q$ collides with the wall, $Q$ reverses direction. Because the collision is perfectly elastic, $Q$’s speed remains the same while its direction changes. Body $Q$ moves toward $P$.\\[2.0em]
    \textbf{(3)} After $Q$ collides again with $P$, $Q$ starts to move toward the wall, continuing the cycle.
  \end{minipage}
  \caption{Schematic sequence of collisions between two bodies and a wall.}
  \label{fig:collision_sequence}
\end{figure}

Before proceeding to the experiment, let us first briefly review the theory behind why the number of collisions related to $\pi$. For more details, see \cite{Galperin, Sanderson}.
Two bodies, body $P$ of mass $M$
and body $Q$ of mass $m$ ($M \ge m$), move along a straight line.  
Body $Q$ lies between the rigid wall and body $P$.
The positive $x$-axis is taken to point away from the wall (to the right in Figure~\ref{fig:collision_sequence}).  
All collisions are perfectly elastic, i.e., we neglect friction, air resistance, and any energy loss due to material deformation, sound, or vibration.
The collision sequence continues until both bodies leave the wall with no further collisions.

In all stages of the collision sequence, we denote the velocities of bodies $P$ and $Q$ by $v_1$ and $v_2$, respectively.
Since all collisions in the sequence are perfectly elastic, both momentum and kinetic energy are conserved at every stage of the collision process. Therefore,
\begin{eqnarray}
M v_1 + m v_2 &=& P_0 \label{eq:momentumconservationv}\\
\frac12 M v_1^{2} + \frac12 m v_2^{2} &=& E_0, 
\label{eq:energyconservationv}
\end{eqnarray}
where $P_0$ and $E_0$ are total momentum and total kinetic energy of the system, respectively.

\begin{figure}[th]
\centering
\begin{minipage}[t]{0.49\columnwidth}
    \centering
    \includegraphics[width=0.9\columnwidth]{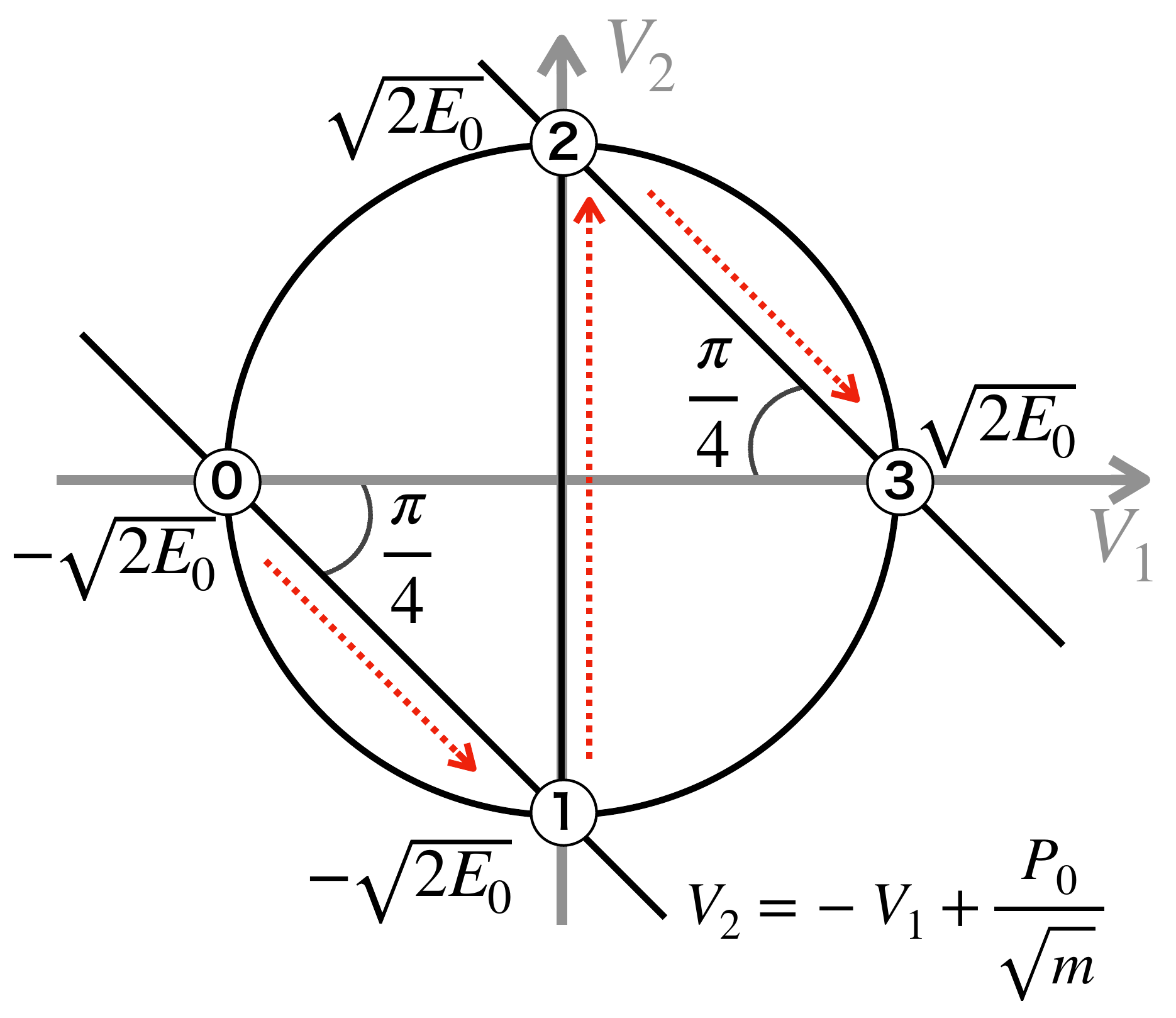}
    \hspace*{-2.5cm}\caption{Case 1 ($m\!:\!M=1\!:\!1$): Motion trajectory in the $V_{1}$–$V_{2}$ plane.}
    \label{fig:phasecircle1}
\end{minipage}
\begin{minipage}[t]{0.49\columnwidth}
    \centering
    \includegraphics[width=0.9\columnwidth]{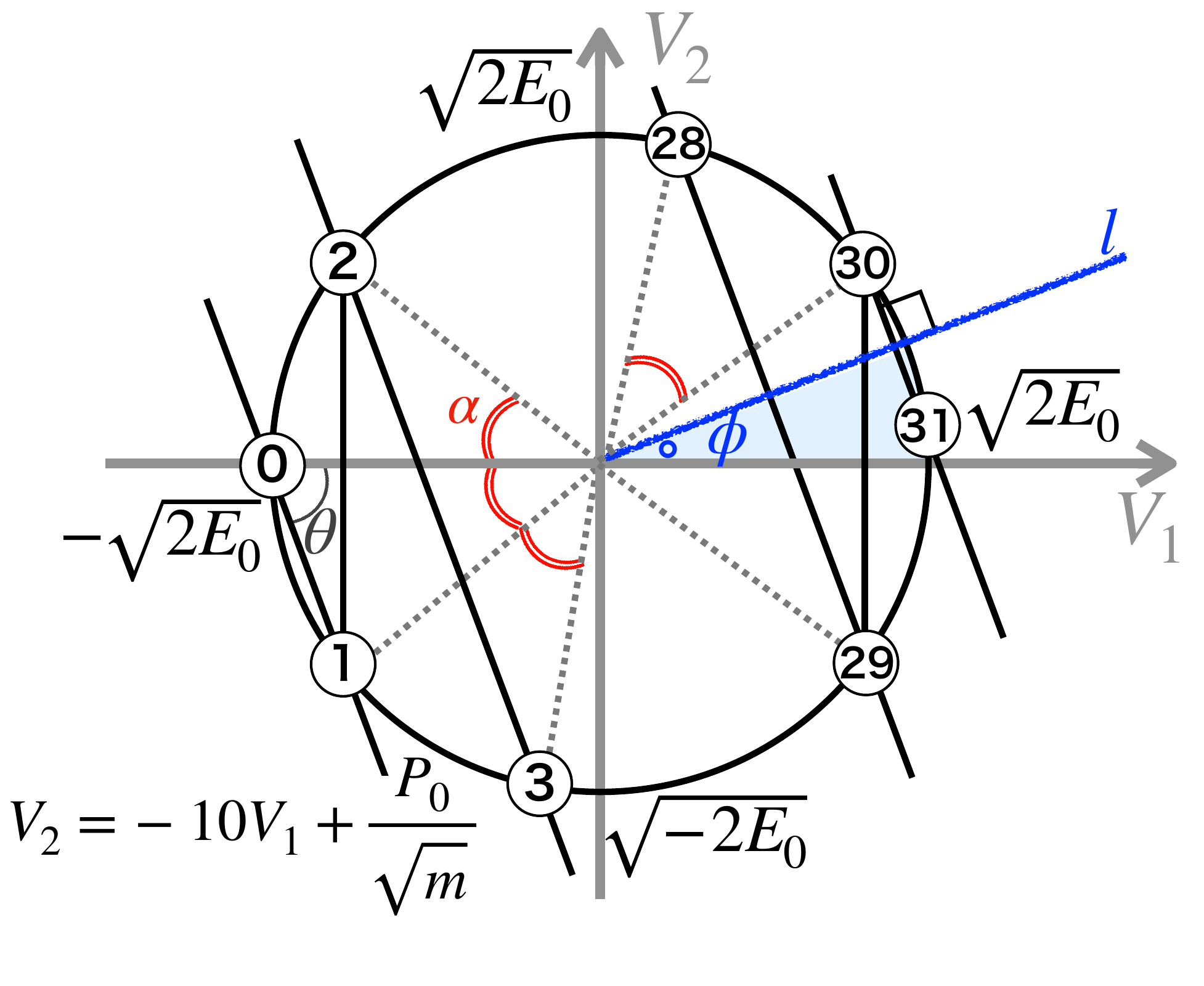}
    \hspace*{-2.5cm}\caption{Same as Fig.~\ref{fig:phasecircle1}, but for Case 2 ($m\!:\!M=1\!:\!100$).}
    \label{fig:phasecircle2}
\end{minipage}
\end{figure}

To make the underlying geometric structure clearer, we rescale the velocity variables
\begin{eqnarray}
V_1 = \sqrt{M}\,v_1, \quad 
V_2 = \sqrt{m}\,v_2.
\label{eq:V12def}
\end{eqnarray}
Replacing the conservation laws eq.~(\ref{eq:momentumconservationv}) and eq.~(\ref{eq:energyconservationv}) in terms of the rescaled velocities of eq.~(\ref{eq:V12def}) gives
\begin{eqnarray}
V_2 &=& -\sqrt{\frac{M}{m}}\;V_1 \;+\; \frac{P_0}{\sqrt{m}}
\label{eq:scaledmomentum}\\
V_1^{2} + V_2^{2} &=& 2E_0.
\label{eq:scaledenergy}
\end{eqnarray}
These equations describe, in the $(V_1,V_2)$ plane, a straight line eq.~(\ref{eq:scaledmomentum}) and a circle eq.~(\ref{eq:scaledenergy}), respectivelly.
For example, the phase‐space diagrams in the $(V_1,V_2)$ plane for the mass ratios $m:M=1:1$ and $m:M=1:100$ are shown in Figure~\ref{fig:phasecircle1} and ~\ref{fig:phasecircle2}.

Thus, as long as momentum and kinetic energy are conserved, the state of the system lies at the intersection of a straight line and a circle in the $(V_1,V_2)$ plane.
Since each point on the plane represents the system’s state immediately after a collision, we call it a state point.
A $P$–$Q$ collision moves the point from one intersection of the chord to the other, while a $Q$–wall collision reflects it vertically, $(V_1, V_2) \to (V_1, -V_2)$, both illustrated by the red dashed arrows in Figure~\ref{fig:phasecircle1}.
In this way, each collision advances the state around the circle, stepping from one discrete point to the next.
We will now examine in detail the cases $m:M=1{:}1$, $1{:}100$, and $1{:}100^{n}$.

%-----------------------------------------------------------------
\subsection{Illustrative Examples}

\subsection*{Case 1 --- $m:M = 1:1$ (Figure~\ref{fig:phasecircle1})}

We first consider the simplest case $M = m$.

\begin{enumerate}[label=(\arabic*), start=0]
  \item Initial state (state point~(0)): We choose the initial condition so that body $P$ moves toward the wall with speed $u>0$ ($v_{1}=-u$), while body $Q$ is at rest ($v_{2}=0$).  Hence
        \[
          V^{(0)}_{1} = \sqrt{M}\,v_{1} = -\sqrt{M}\,u,\quad
          V^{(0)}_{2} = \sqrt{m}\,v_{2} = 0.
        \]
        This is the leftmost point $(0)$ on the circle.

  \item State point~(1): State after the first collision ($P\to Q$). Body $P$ moves leftward with $v_{1}=-u$ and collides elastically with the stationary $Q$. 
  Because the collision is perfectly elastic and the masses are equal, $P$ comes to rest and $Q$ takes all the momentum, resulting in $v_{1}=0$ and $v_{2}=-u$. The scaled velocities become
  \[
    V^{(1)}_{1}=0,\quad 
    V^{(1)}_{2}=-\sqrt{m}\,u.
  \]
    In the $(V_{1},V_{2})$ plane, the state point moves along the momentum‐conservation line toward the lower right, now reaching the other intersection $(1)$ with the energy‐conservation circle.
    The momentum‐conservation line has slope $-\sqrt{M/m} = -1$, and thus makes a $\pi/4$ angle with the $V_{1}$‐axis.

  \item State point~(2): State reached after the second collision $Q\to $wall. $Q$'s speed remains $u$ but its direction reverses, resulting in $v_{2}=+u$ while $v_{1}=0$. 
      The scaled velocities then become
      \[
        V^{(2)}_{1}=0,\quad V^{(2)}_{2}=+\sqrt{m}\,u.
      \]
      This reflects the state vertically across the $V_{1}$-axis, moving it to the corresponding intersection $(2)$ above the $V_1$ axis on the circle.

    \item State point (3):  
    State reached after the third collision $Q\to P$. In this perfectly elastic collision, $Q$ transfers all its momentum back to $P$, resulting in $v_{1}=+u$ and $v_{2}=0$. Thus the corresponding scaled velocities are
      \[
        V^{(3)}_{1}=+\sqrt{M}\,u,\quad V^{(3)}_{2}=0.
      \]
      The state arrives at the rightmost point $(3)$ on the circle.
\end{enumerate}

After these three collisions, the velocities satisfy $0< v_{2}< v_{1}$, 
so body P moves faster than body Q, and no further collisions occur.
Thus, during the collision sequence the system’s state visits four velocity configurations, which correspond to four points on the circle. 
Each collision advances the state from one point to the next, so the total number of collisions $k$ is one less than the number of state points.
There are four points on the circle, thus the number of collisions is 
\[
k = 4 - 1 = 3.
\]
In this way, the first digit of $\pi$, 3, emerges as the number of collisions.

%-----------------------------------------------------------------
\subsection*{Case 2 --- \ $m:M = 1:100$ \ (Figure~\ref{fig:phasecircle2})}

%--------------------------------------------------------------
% Case 2  :  m : M  =  1 : 100
%--------------------------------------------------------------

The next simplest case is $m:M = 1:100$, which also corresponds to the experimental setup considered in this study.
In this case, the system’s state visits the following points on the energy‐conservation circle.

\begin{enumerate}[label=(\arabic*),start=0]

  \item State point~(0): 
As in Case 1, body $P$ is released toward the left with speed $u>0$, while $Q$ is initially at rest, thus 
$v_{1}=-u$, $v_{2}=0$. 
Scaled velocities are 
        \[
          V^{(0)}_{1}=\sqrt{M}\,v_{1}=-10\sqrt{m}\,u,
          \qquad
          V^{(0)}_{2}=\sqrt{m}\,v_{2}=0 .
        \]

  \item State point~(1): State reached after the first elastic collision $P\to Q$. 
        Denoting the pre-collision velocities by $v_{1},v_{2}$ and the post-collision ones by $v'_{1},v'_{2}$, conservation of momentum is
\[
  M\,v_{1} + m\,v_{2} = M\,v_{1}' + m\,v_{2}'.
\]
A perfectly elastic collision has a coefficient of restitution
\[
  -\frac{v_{2}' - v_{1}'}{v_{2} - v_{1}} = 1,
\]
which, of course, holds when both momentum and energy are conserved.
Solving these two equations for an initial state $v_{1}=-u$, $v_{2}=0$ yields
\begin{eqnarray}
&v_{1}' = &-\frac{99}{101}\,u,\quad v_{2}' = -\frac{200}{101}\,u \nonumber\\
&\Longrightarrow& 
V^{(1)}_{1}=-10\sqrt{m}\,\frac{99}{101}\,u,
          \qquad
          V^{(1)}_{2}=-\sqrt{m}\,\frac{200}{101}\,u . \nonumber
\end{eqnarray}
    The state moves along the momentum-conservation line to the lower-right intersection (1) of the circle.
    The momentum‐conservation line has slope $-\tan\theta = -\sqrt{M/m} = -10$ (See Figure~\ref{fig:phasecircle2} for $\theta$).

  \item State point~(2):  
    State reached after the elastic collision of body $Q$ with the rigid wall.  $Q$'s speed remains $(200/101)u$ but the direction reverses ($v'_{2}=+200/101\,u$, while $v'_{1}$ unchanged).
        Now the scaled velocities are
        \[
          V^{(2)}_{1}=-10\sqrt{m}\,\frac{99}{101}\,u,
          \qquad
          V^{(2)}_{2}=+\sqrt{m}\,\frac{200}{101}\,u .
        \]
        Geometrically the state reflects vertically across the $V_{1}$–axis to the upper intersection (2) of the line and circle.
\end{enumerate}

The angle $\alpha$ between the state at point (2) and the $V_{1}$-axis can be written as
\begin{eqnarray}
  \alpha
  = \arccos\left(
      \frac{|V_{1}^{(2)}|}
           {\sqrt{\bigl(V_{1}^{(2)}\bigr)^{2} + \bigl(V_{2}^{(2)}\bigr)^{2}}}
    \right)
  %= \arccos\Bigl(\,\frac{990\sqrt{m}u}{1010\sqrt{m}\,u}\Bigr)
  = \arccos\Bigl(\frac{99}{101}\Bigr)
  \simeq 0.199.
  \label{eq:alpha1_100}
\end{eqnarray} 
After state point (2), the sequence of slanted jumps (for the $Q-P$ collisions) and vertical reflections (for the wall collisions) continues in the same way. We omit the remaining steps here.

So far, we have described the motion using simple kinematic arguments.  Now we turn to the geometric structure of the state points and their connections on the circle in the $(V_1, V_2)$ plane.
Starting from the leftmost point $(0)$ on the circle, the sequence of states proceeds
right-down, up, right-down, up, … so that the point is placed on the
circle at equal angular intervals of size $\alpha$.  
Unlike Case 1, when $\alpha$ does not divide $2\pi$ exactly, the counting requires a little care.
Because every pair of collisions (slanted jump $+$ vertical reflection)
produces one point in the lower half-circle and the next point in the upper
half-circle, it is sufficient to count the points that appear in the upper half.  
From the leftmost point (0) to the rightmost side, the even-numbered upper
points $(2),(4),(6),\ldots$ advance by $\alpha$ each time.  
Since $\alpha \simeq  0.199$ (from eq.~\ref{eq:alpha1_100}), we find there are $15$ such points in the upper half of the circle.  
The same $15$ points appear in the lower half of the circle, thus so far, together with the initial point $(0)$, there are
\begin{eqnarray}
1+15\times2=31
\label{eq:counting1}
\end{eqnarray}
points.

Since the final collision may be either a vertical reflection or a slanted jump, it remains unclear whether the sequence ends with the vertical reflection. 
Thus, let us examine in detail the condition for the sequence of collisions to end.
When no further collisions occur, the two bodies move away from the wall, and body $P$ must be faster than body $Q$.
This requires $0 < v_2 < v_1$.
In terms of the scaled velocities $V_1=\sqrt{M}v_1$ and $V_2=\sqrt{m}v_2$, this condition becomes
\[
0<V_2<\sqrt{\frac{m}{M}}\,V_1.
\]
Geometrically, this corresponds to the region bounded by the line $l: V_2 = \sqrt{\frac{m}{M}}\,V_1$, which passes through the origin and is perpendicular to the momentum conservation lines, and by the $V_1$ axis. 
It is shown as the lightly shaded blue region in Figure~\ref{fig:phasecircle2}. 
In other words, once the state point enters this region, no further collisions can occur.
The line $l$ is orthogonal to the momentum conservation line, so we have  $\tan\theta\,\tan\phi=1$ where $\phi = \arctan\sqrt{\frac{m}{M}}$ denotes the slope angle of line $l$.
For the present mass ratio $m:M=1:100$, the boundary becomes 
$l:\,V_2=\frac{1}{10}V_1$, and the line $l$ makes an angle 
$\phi=\arctan(1/10)\simeq 0.0997$ with the $V_1$ axis. 
The 15-th upper half point $(30)$ obtained by the last vertical reflection has an
angle measured from the $V_1$ axis of 
$\pi-15\,\alpha=\pi-15\,\arccos(99/101)\simeq 0.1515$. 
Since $0.1515>0.0997$, point $(30)$ lies above the line $l$ and therefore does not satisfy 
the stopping condition. 
One more slanted jump from point $(30)$ to the lower right is required. 
As a result, the final point $(31)$ must lie on the circle. 
The total number of state points on the circle is 
$1+15\times2+1=32$.
The number of collisions, which is one fewer than the number of state points, is
\[
k = 32 - 1 = 31 .
\]
Thus, for the mass ratio $m:M=1:100$, the system has 31
collisions, which yields the first decimal digit of $\pi$.

%-----------------------------------------------------------------
%-----------------------------------------------------------------
\subsection*{Case 3 --- $m:M = 1:100^{\,n}$}

As in Cases~1 and~2, we now extend to the general mass ratio $1:100^{\,n}$ and examine explicitly the first few state points of the system’s time evolution.
Investigating the general case offers a clearer perspective on how the number of collisions relates to $\pi$.

\begin{enumerate}[label=(\arabic*), start=0]
  \item State point~(0):  
        Body $P$ (mass $M=100^{\,n}m$) moves toward the wall at speed $u>0$ ($v_{1}=-u$), while body $Q$ is at rest ($v_{2}=0$);  
        \[
          V^{(0)}_{1} = -10^{\,n}\sqrt{m}\,u,\quad
          V^{(0)}_{2} = 0,
        \]
        which corresponds to the leftmost point $(0)$ on the circle in  Figure~\ref{fig:phasecircle2}. 
\[
  V^{(0)}_{1} \;=\;\sqrt{M}\,v_{1} = -\,10^{n}\sqrt{m}\,u,\quad
  V^{(0)}_{2} \;=\;\sqrt{m}\,v_{2} = 0,
\]
corresponding to the leftmost point (0) on the circle in Figure 3.

  \item State point~(1):  
  After the first collision ($P\to Q$), momentum conservation
  \begin{equation}
    100^{n}(-u)=100^{n}v'_{1}+v'_{2}
    \label{eq:nmomentumconservation}
  \end{equation}
  and the fact that the coefficient of restitution is $1$
\begin{equation}
  -\frac{v_2' - v_1'}{v_2 - v_1} = 1
  \label{eq:nrestitution}
\end{equation}
yield two simultaneous equations. 
From these two equations,
  \begin{equation}
    v'_{1}=-\frac{100^n-1}{100^n+1}\,u,\quad
    v'_{2}=-\frac{2\cdot100^n}{100^n+1}\,u.
  \end{equation}
The state moves to the lower-right intersection (1) of the momentum conservation line and the circle.
  The scaled velocities are now
  \[
\hspace*{-4em} 
    V^{(1)}_{1} = \sqrt{M}\,v'_{1}
            = -\frac{100^n-1}{100^n+1}\,10^n\sqrt{m}\,u,\quad
    V^{(1)}_{2} = \sqrt{m}\,v'_{2}
            = -\frac{2\cdot100^n}{100^n+1}\,\sqrt{m}\,u,
  \]
  marking the lower‐right intersection $(1)$ on the circle.

  \item State point~(2):  
        After $Q$ hits the wall elastically, its speed is unchanged but its direction reverses ($v''_{2}=+2\cdot 100^{\,n}u/(100^{\,n}+1)$, while $v''_{1}=v'_{1}$), and 
        \[
          V^{(2)}_{1} = -\frac{100^{\,n}-1}{100^{\,n}+1}\,10^{\,n}\sqrt{m}\,u,\quad
          V^{(2)}_{2} = +\frac{2\cdot 100^{\,n}}{100^{\,n}+1}\sqrt{m}\,u,
        \]
        which is the vertical reflection across the $V_{1}$-axis.
\end{enumerate}

As shown in Fig.~\ref{fig:phasecircle2}, the angle between the $V_{1}$-axis and state point $(2)$ satisfies
\begin{eqnarray}
  \alpha = \arccos\left(
      \frac{|V_{1}^{(2)}|}
           {\sqrt{\bigl(V_{1}^{(2)}\bigr)^{2} + \bigl(V_{2}^{(2)}\bigr)^{2}}}\right)
           = \arccos\left(\frac{100^n - 1}{100^n + 1}\right),
  \label{eq:alpha}
\end{eqnarray}
where $V_1^{(2)}$ and $V_2^{(2)}$ are the scaled velocities at point (2).
Starting from point $(0)$, the motion proceeds counterclockwise along the lower half of the circle, and the wall reflection sends the state to the upper half.
As long as we consider only the upper half of the circle, the relevant angular range is $\pi$, not the full $2\pi$.
We will focus on the upper half of the circle, where the system is located at the even-numbered state points $(2), (4), (6) \cdots$, each separated by an angular step of $\alpha$.
We define $j_{\max}$ as the largest integer $j$ satisfying $j\alpha<\pi$, then
\[
  j_{\max} = \left\lfloor \frac{\pi}{\alpha} \right\rfloor,
\]
  where $\lfloor x\rfloor$ denotes the floor function (the greatest integer not exceeding $x$).
Here, $j_{\max}$ corresponds to the rightmost state point on the upper half of the circle reached by taking repeated steps of size $\alpha$ around the circle.
Then there are $j_{\max}$ state points on the upper half of the circle, namely $(2), (4), \cdots, (2j_{\max})$.  
Each of those $j_{\max}$ points on the upper half has a corresponding point on the lower half, for a total of $2j_{\max}$ states.  
Including the leftmost state point (0) increases the count to $2j_{\max}+1$. 
As discussed in Case~2, whether the collision sequence terminates at the even-numbered state point $(2j_{\max})$, created by the final vertical reflection from the lower to the upper half of the circle, depends on whether that point $(2j_{\max})$ lies inside the stopping region.  
Specifically, if the angle between $(2j_{\max})$ and the $V_1$-axis is smaller than $\phi = \arctan\sqrt{m/M}$, then $(2j_{\max})$ is the final state,  
while if the angle is larger than $\phi$, the situation reduces to Case~3: one additional slanted jump to the lower right is required, and the point $(2j_{\max}) + 1$ becomes the final state.
Hence, the total number of state points on the circle $N$ is either $2j_{\max}+1$ or $2j_{\max}+2$.
Since the number of collisions $k$ is one fewer than the number of state points, we have
\begin{eqnarray}
k=N-1=
\left\{
\begin{array}{ll}
2\,j_{\max} =2\,\lfloor \pi/\alpha \rfloor      & \mathrm{if}\; N = 2\,j_{\max}+1,\\[6pt]
2\,j_{\max}+1 =2\,\lfloor \pi/\alpha \rfloor + 1    & \mathrm{if}\; N = 2\,j_{\max}+2.
 \label{eq:conclusion1}
\end{array}
\right.
\end{eqnarray}

Let us now consider the case of large $n$. 
The expression for $\cos\alpha$ can be written as
\[
  \cos\alpha = \frac{100^n - 1}{100^n + 1} = 1 - \varepsilon,
  \quad
  \varepsilon = \frac{2}{100^n+1}.
\]
In the large-$n$ limit, where $\varepsilon$ is small, we can approximate $\arccos(1 - \varepsilon)$ as $\sqrt{2\varepsilon}$, thus
\[
  \alpha =\arccos{\left(\frac{100^n - 1}{100^n + 1} \right)} 
  \simeq \sqrt{\frac{4}{100^n+1}} \simeq 2\times10^{-n}.
\]
From equation~(\ref{eq:conclusion1}), the total number of collisions in the large $n$ limit can be written as
\begin{eqnarray}
  k \;=\;
  \left\{
        \begin{array}{l}
         2 \left\lfloor \frac{\pi}{2} 10^{n} \right\rfloor, 
           \mathrm{if}\; N = 2\,j_{\max} + 1, \\[6pt]
         2 \left\lfloor \frac{\pi}{2} 10^{n} \right\rfloor + 1, 
           \mathrm{if}\; N = 2\,j_{\max} + 2
       \end{array}
       \right.
       \;\longrightarrow\; 10^{n}\,\pi.
\end{eqnarray}
Thus, for large $n$, the collision count $k$ equals $\pi$ up to its $n$-th decimal place. However, note that the large $n$ limit is not applicable to the present study, since it considers the case $n = 1$.

%-----------------------------------------------------------------

\section{Setup of the experiment}
\label{sec:experiment-setup}

\begin{table}
\caption{Setup of the experiment. The definitions of length between body-body are shown in Figure \ref{fig:overviewdiagram}.}
\footnotesize
\begin{center}
\begin{tabular}{@{}lc} 
\br
Body-Body& Length [cm]\\
\mr
H-L (original position)&0\\
H-L (when H is released)&9\\
L-W (original position)&7\\
\mr \mr
mass of H & 1050 g\\
mass of L & 10.5 g\\
\br
\end{tabular}
\end{center}
\label{tab:setup}
\end{table}
\normalsize
\begin{figure}[h]
\begin{center}
\includegraphics[width=8cm]{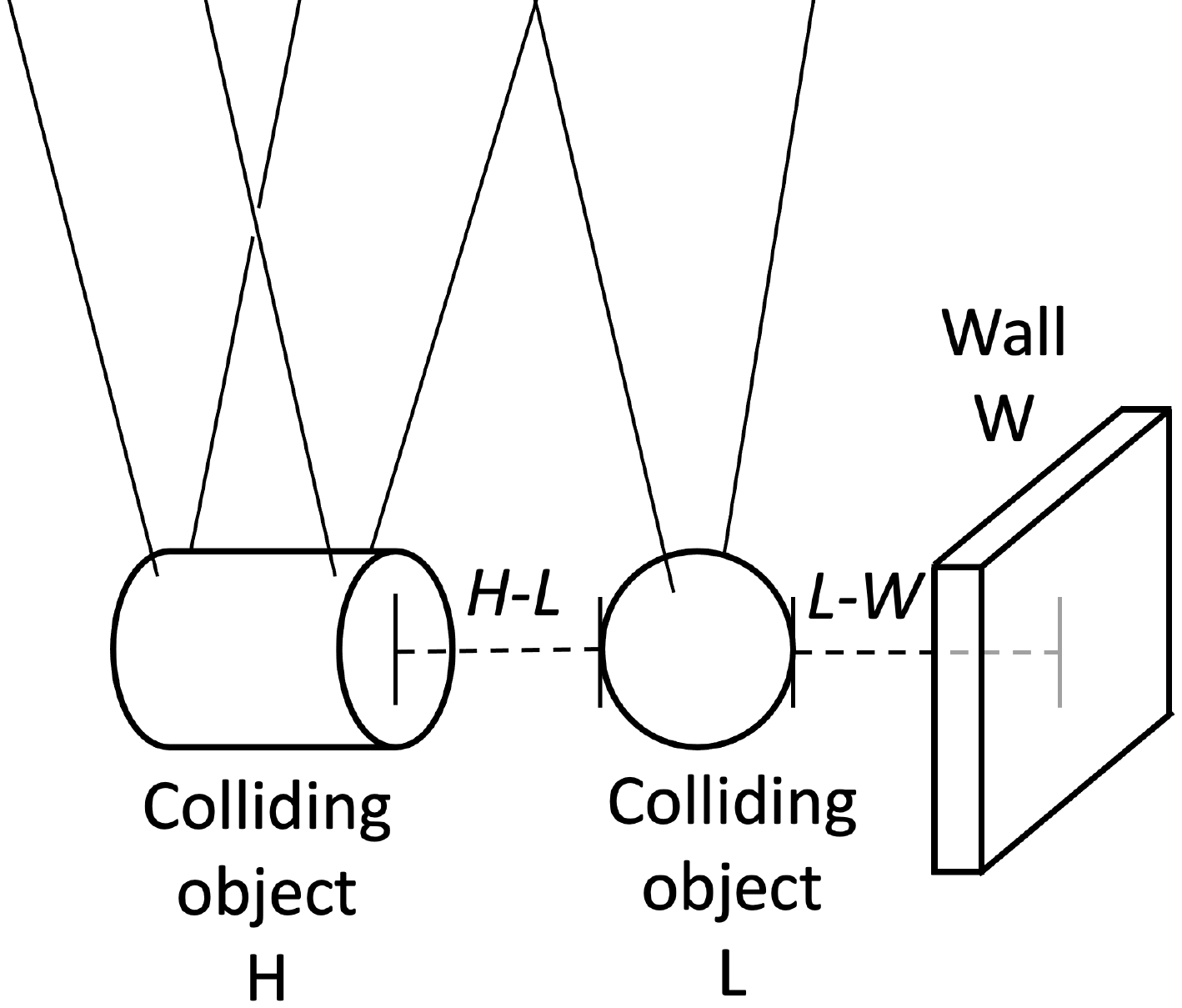}
\caption{Overview diagram of experimental apparatus.}
\label{fig:overviewdiagram}
\end{center}
\end{figure}
\begin{figure}[h]
\begin{minipage}[b]{0.49\columnwidth}
    \centering
    \includegraphics[width=0.9\columnwidth]{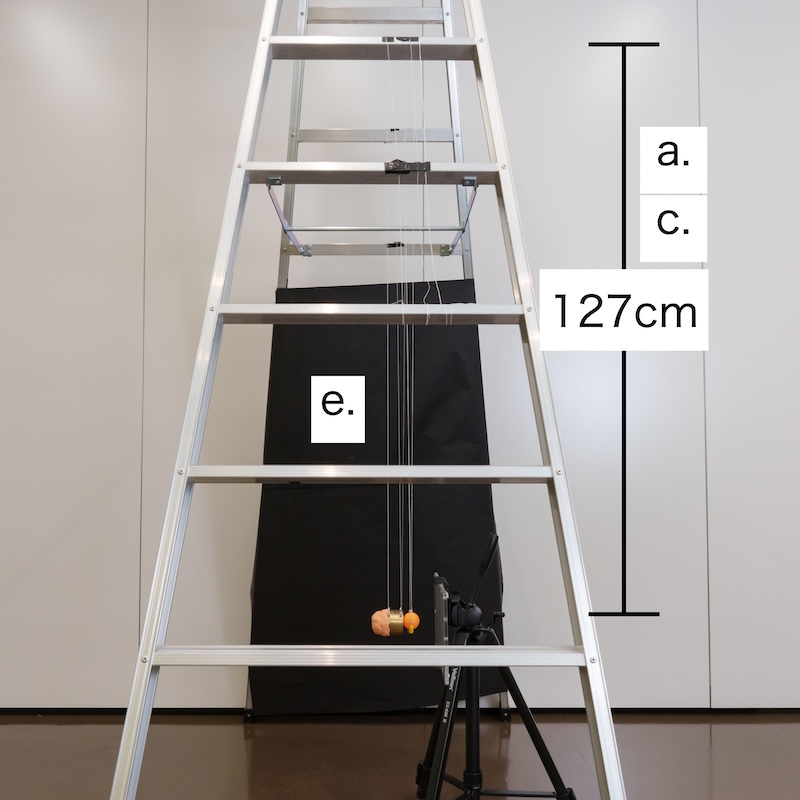}
    %\caption{left}
\end{minipage}
\begin{minipage}[b]{0.49\columnwidth}
    \centering
    \includegraphics[width=0.9\columnwidth]{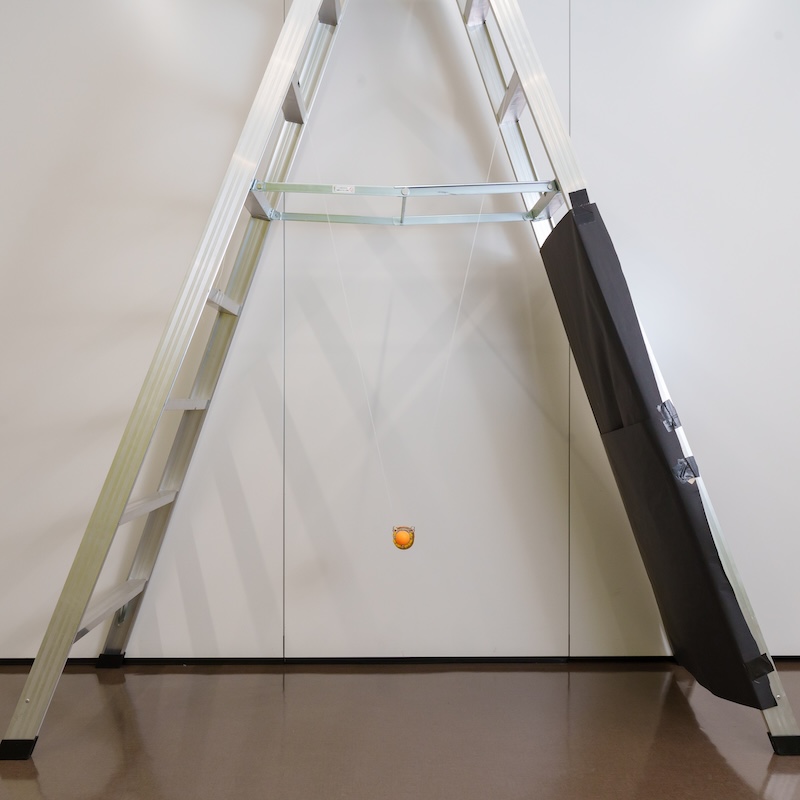}
    %\caption{right}
\end{minipage}
    \centering
    \caption{Photographs of the whole experimental apparatus. In the left figure, the wall has been removed so that the colliding bodies can be photographed.}
    \label{fig:overview23}
\end{figure}
\begin{figure}[h]
\centering
    \includegraphics[width=8cm]{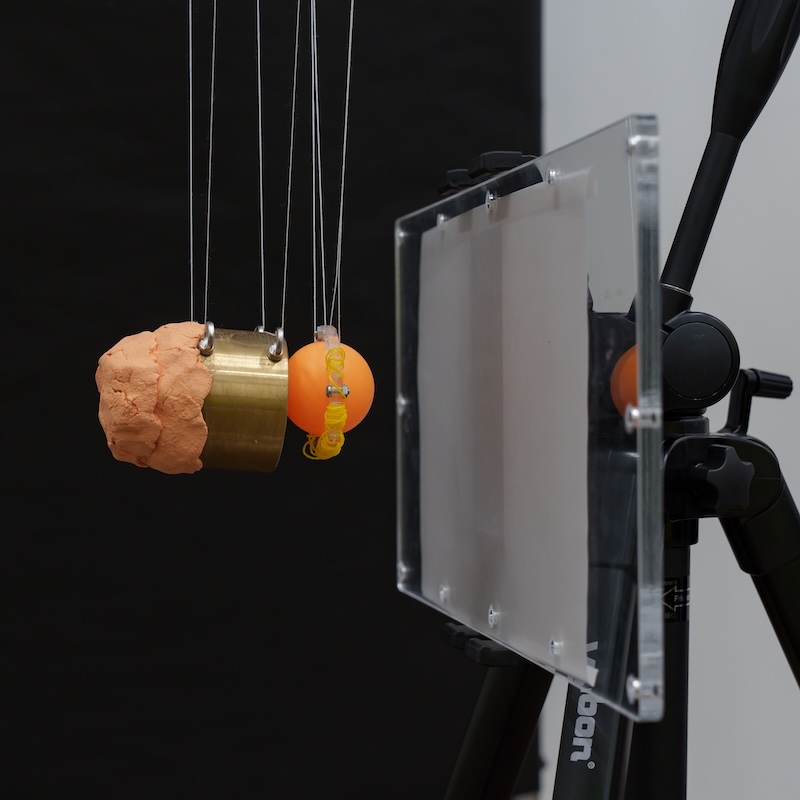}
    \caption{A close-up of the experimental apparatus.}
    \label{fig:closeup1}
\end{figure}
\begin{figure}[h]
\centering
\begin{minipage}[b]{0.49\columnwidth}
    \centering
    \includegraphics[width=0.9\columnwidth]{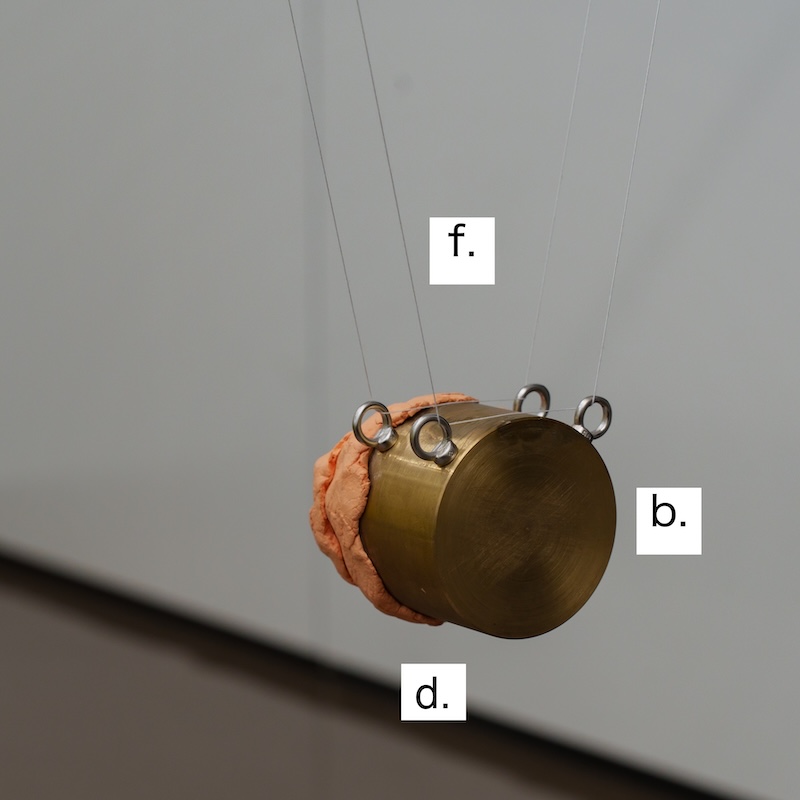}
    %\caption{left}
\end{minipage}
\begin{minipage}[b]{0.49\columnwidth}
    \centering
    \includegraphics[width=0.9\columnwidth]{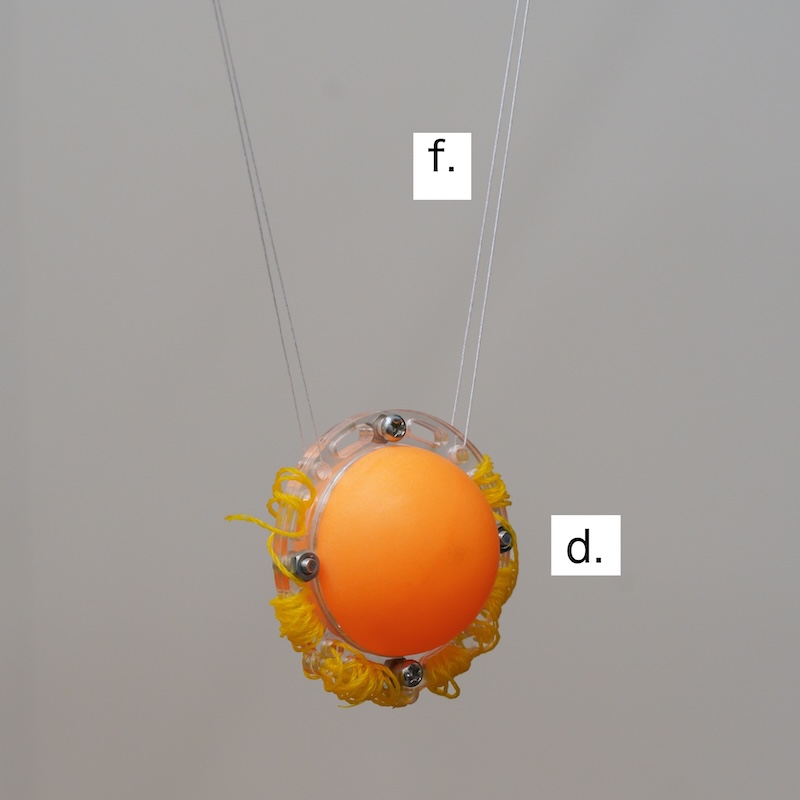}
    %\caption{right}
\end{minipage}
    \centering
    \caption{Close view of the colliding bodies H (left) and L (right).}
    \label{fig:closeup23}
\end{figure}

In an experiment involving repeated collisions between two colliding bodies and a wall, the major possible sources of energy loss in the system are friction between the colliding bodies and the floor, the inelasticity of the collision, and the air resistance acting on the colliding bodies. In addition, since the colliding bodies are not actually point masses, it could lose energy due to rotation as a rigid body. After some preparatory experiments, we found that the most significant energy loss is friction acting between the mass point and the floor. Therefore, we designed a suspended apparatus, as shown in Figure \ref{fig:overviewdiagram}, to reduce all sources of unnecessary friction.
In Figure~\ref{fig:overviewdiagram}, the heavier (lighter) colliding body is represented as H (L).
Photographs of the actual experimental apparatus are shown in Figures \ref{fig:overview23}-\ref{fig:closeup23}.

Materials used to reduce the energy loss of the suspended devices are listed below. The symbols a.-f. correspond to those in Figures 2-4.

\begin{itemize}
\item[a.] To eliminate friction between the colliding bodies and the floor, the colliding bodies are suspended by a string in air.
\item[b.] Instead of making both colliding bodies spherical, it is better to make one of the colliding bodies flat to avoid blurring of the colliding points.
\item[c.] The string suspending the colliding bodies should be long enough. If the string is short, the heavier body moves along a circular arc rather than a straight line, disrupting the alignment of the collisions.
%\item[c.] The string suspending the colliding bodies should be long enough. If the string is short, the motion of the colliding body is no longer parallel to the floor, resulting in a circular arc trajectory of the transducer.
\item[d.] In this experiment, we used a brass cylinder, a table tennis ball, and an acrylic plate.
Because it was difficult to achieve precise mass adjustment through machining alone, the brass cylinder was intentionally fabricated slightly underweight, and modeling clay was added to the back of the cylinder to reach the target mass ratio of 1:100. 
This caused a minor asymmetry in the geometry of the brass cylinder, potentially shifting the center of mass away from the ideal collision axis. However, since collisions occurred stably at the flat face of the brass cylinder and the observed number of collisions matched the theoretical prediction (31), we concluded that this effect was not significant for the level of precision targeted in this experiment.
While acrylic is not ideal as a wall material in terms of rigidity and elasticity, it was sufficient for achieving 31 collisions, corresponding to the first decimal digit of $\pi$. For higher-precision experiments, a more rigid wall material may be preferable.

\item[e.] The material of the string used for suspension must also be devised so that energy loss will not be caused by the string stretching and contracting. In this study, piano wire, fishing line, octopus thread (kite string), and fine sewing thread were tested. We chose fine sewing thread because the collision was the most stable with it.
\item[f.] To ensure stable suspension, each body was hung using two to four threads. Especially, in the case of the lighter body (table tennis ball), a plastic ring was used to provide a rigid mounting point for the threads. This design allowed the ball to be suspended at two points symmetrically, which prevented undesirable rotation during collisions. The mass of the lighter body, as used in the analysis, includes the ring.
%If the colliding body is suspended by only one thread, the direction of motion becomes unstable and the axis of collision shifts as the collision is repeated several times. Therefore, the axis of the colliding bodies were fixed by suspending them at two or more points with two to four threads. %The detailed setup values are listed in Table \ref{tab:setup}.
\end{itemize}

\section{The experiment and result}
\label{sec:experiment-result}

This section describes the experiment%\footnote{For a demonstration of the experiment with brief explanations, see \href{https://youtu.be/jjWg_GH_slY}{this video} \cite{KIN1}. For raw experiment footage, see \href{https://youtu.be/0sKfW7Sagog}{this video} \cite{KIN2}.}
 performed based on the setup described in Section \ref{sec:experiment-setup}. The experiment was conducted a total of 30 times using the following procedure.
\begin{enumerate}
%\item Based on the experimental setup shown in Figure \ref{fig:overviewdiagram}, the colliding body H was placed 7 cm away from the colliding body L, then body H was gently released, and it collided with body L, lastly body L collided with the acrylic plate B.
\item Based on the experimental setup shown in Figure~\ref{fig:overviewdiagram}, the colliding body H was placed 7~cm away from the colliding body L, then body H was gently released, and it collided with body L, lastly body L collided with the acrylic plate B.
While Section~\ref{sec:theory} provides a theoretical explanation of the system, the body-to-body and body-to-wall distances are not part of the idealized model, as they do not affect the total number of collisions in the absence of friction and other dissipative forces. However, in practice, these effects cannot be ignored, and if the positioning is not appropriate, the number of collisions may fall short of the theoretical value of 31. Therefore, to ensure reproducibility and minimize experimental deviations, we adopted the setup values that most consistently produced the expected outcome. These values are listed in Table~\ref{tab:setup}.

\item The scene was recorded with a slow-motion camera, and the number of collisions was counted by reviewing the video.
\end{enumerate}

Table \ref{tab:result} shows the number of collisions and other information counted in the experiment. 
Although the theoretical value for the number of collisions is 31, this value is derived under idealized conditions where friction, rotational effects, and misalignment are entirely negligible. In practice, however, even with careful design, some energy dissipation is inevitable. As a result, some trials yield fewer collisions than the theoretical value. To address this, we refined the experimental setup through repeated adjustments, selecting conditions that most consistently produced results close to 31. For this reason, we report both the maximum number of collisions observed—which corresponds to the theoretical prediction—and the mean, standard deviation and the minimum number of collisions over 30 trials, which reflect the reproducibility and stability of the optimized setup.
As can be seen from Table \ref{tab:result}, 
the maximum observed number of collisions matched the theoretical value (31). 
The standard deviation of 0.80 reflects the stability of the setup, indicating reliable experimental conditions.
The average number of collisions for the 30 measurements is 30.57, also indicating that this setting can reproduce the theoretical value relatively stably.

\begin{table}
\caption{Result of the experiment. The experiment was conducted 30 times.}
\footnotesize
\begin{center}
\begin{tabular}{@{}lc} 
\br
 & Number of collisions\\
\mr
Maximum value &31\\
Minimum value &28\\
Mean value &30.57\\
Standard deviation &0.80\\
\br
\end{tabular}
\label{tab:result}
\end{center}
\end{table}
\normalsize

\section{Conclusion and future prospects}
\label{sec:conclusion}
In this paper, $\pi$ is obtained to one decimal place not by mathematical means, but via a physical experiment.
Although it has been theoretically shown that $\pi$ can be estimated in an ideal environment where friction and air resistance are negligible, in reality, they cannot be ignored, making an accurate estimation of $\pi$ generally difficult.
This study will serve as a good teaching tool to consider the difference between an ideal environment (an environment without friction or air resistance) and real conditions.

%The results of the preparatory experiment indicated that friction between the colliding bodies and the floor is a major problem. Therefore, we devised an experimental apparatus to eliminate friction by suspending the bodies in air.As a result, when the mass ratio of the two bodies was $1:100$, we succeeded in measuring 31 collisions, which is the theoretical value for the ideal situation where the energy loss can be ignored.This corresponds to 3.1, representing $\pi$ up to the tenth decimal place.Future work will focus on increasing the mass ratio to $1:10,000$ to measure $\pi$ to the hundredth decimal place; 3.14.This will require either significantly heavier masses or innovative methods to reduce string-induced energy losses.If the same apparatus as this experiment is used, the heavier body must be significantly heavier, like $\sim 100$ kg, and precautions must be taken to ensure the safety of the experimenters.If the mass ratio is adjusted by making the light body much lighter, the light body will be pulled by the tension of the string, causing energy loss due to the expansion and contraction of the string. Therefore, we need to improve our experimental apparatus to achieve the desired mass ratio. In any case, the next step, the measurement of $\pi$ for a mass ratio of 1:10000, will not be easy.

The results of the preparatory experiment indicated that friction between the
colliding bodies and the floor is a major problem. Therefore, we devised an experimental apparatus to eliminate friction by suspending the bodies in air. As a result, when the mass ratio of the two bodies was 1 : 100, we succeeded in measuring 31 collisions, which is the theoretical value for the ideal situation where the energy loss can be ignored. 
This corresponds to 3.1, representing $\pi$ up to the first decimal place. 

In the theoretical model, the number of collisions is finite due to geometric constraints in the velocity phase space. In contrast, in physical experiments, collisions may terminate either for the same geometric reasons or because energy is gradually lost due to inelasticity or friction. These two mechanisms—geometric and dissipative—can lead to similar outcomes but arise from different principles, and they are easily confused.
In the present experiment, although some energy was inevitably lost during collisions, both bodies retained significant kinetic energy after the final collision. This indicates that the finite number of collisions was not primarily due to dissipation, but rather reflected the geometric structure of the phase space. This distinction helps clarify the physical origin of the observed behavior.

Future work will focus on increasing the mass ratio to 1 : 10,000 to estimate $\pi$ to the hundredth decimal place, 3.14. This will require either significantly heavier masses or innovative methods to reduce string-induced energy losses. If the same apparatus as this experiment is used, the heavier body must be significantly heavier, such as $\sim 100\ \mathrm{kg}$, and precautions must be taken to ensure the safety of the experimenters. If the mass ratio is adjusted by making the light body much lighter, the light body will be pulled by the tension of the string, causing energy loss due to the expansion and contraction of the string. Therefore, we need to improve our experimental apparatus to achieve the desired mass ratio. In any case, the next step, the estimation of $\pi$ for a mass ratio of 1:10000, will not be easy.

\ack
This work was supported by Science Dream Lab, Design and Manufacturing Center, Okayama University of Science. 
KIN is grateful to Tatsuya Ishihara for collaboration in the early stages of this work. 
KIN also would like to thank Shinya Higa for the clear photos and videos, Jun'ya Hori for providing a slow-motion camera, and Nicholas J. Benoit for precious comments to the draft.

\end{document}